\documentclass[12pt,a4paper]{article}
\usepackage{latexsym}
\usepackage{graphicx}
\pdfoutput=1
\usepackage{pdfpages}   
\usepackage{color}
\usepackage{amsmath,amssymb}
\usepackage{xspace,textcomp}
\usepackage{ifthen}
\usepackage{hyperref}
\usepackage{slashed}
\usepackage{multirow}
\usepackage[rightcaption]{sidecap}
\sidecaptionvpos{figure}{c}
\newcommand{\beq}{\begin{equation}}   %
\newcommand{\eeq}{\end{equation}}   %
\newcommand{\be}{\begin{eqnarray}}
\newcommand{\ee}{\end{eqnarray}}

\newcommand{\aMCatNLO}{{\sc a}MC@NLO}

\begin{document}

\thispagestyle{empty}

\begin{center}
\hfill RECAPP-HRI-2014-005
\end{center}

\begin{center}

\vspace{1.7cm}

{\Large\bf{ 
Three photon production to NLO+PS accuracy at the LHC}}

\vspace{1.4cm}

M.\ K.\ Mandal~$^{a,}$\footnote{mandal@hri.res.in}\ ,
\hspace{.25cm} 
Prakash Mathews~$^{b,}$\footnote{prakash.mathews@saha.ac.in}\ ,
\hspace{.25cm} 
V. Ravindran$~^{c,}$\footnote{ravindra@imsc.res.in}\ ,
\hspace{.25cm} 
Satyajit Seth~$^{b,}$\footnote{satyajit.seth@saha.ac.in}
\\[.5cm]

${}^a$ Regional Centre for Accelerator-based Particle Physics\\ 
Harish-Chandra Research Institute, Chhatnag Road, Jhunsi,\\
Allahabad 211 019, India
\\[.5cm]

${}^b$ Saha Institute of Nuclear Physics,\\
 1/AF Bidhan Nagar,\\
  Kolkata 700 064, India
\\[.5cm]

${}^c$The Institute of Mathematical Sciences\\
C.I.T Campus, 4th Cross St, Tharamani Chennai,\\
Tamil Nadu 600 113, India
\\[.5cm]
\end{center}

\vfill

\centerline{\bf Abstract}
\vspace{2 mm}
\begin{quote}
\small
In this paper, we present the next-to-leading order predictions for
three photon production in the Standard Model, matched to the parton
shower using the MC@NLO formalism. We have studied the role of parton shower on various observables and 
shown a selection of results for the $14$ TeV LHC.
\end{quote}

\vfill

\newpage
\section{Introduction}
Wealth of data from the Large Hadron Collider (LHC) and the Tevatron 
involving large number of leptons, gauge bosons and hadrons 
in the final state not only provides ample opportunity to test the predictions 
of the Standard Model (SM), but also constrains various physics scenarios in the beyond standard model (BSM). 
Signatures of BSM are often plagued by the large SM background
and hence careful study of wide variety of SM processes has been underway \cite{wishlist}.
Precise predictions for such SM processes are important as the 
quantum corrections are often comparable to the BSM effects. In addition, they are essential to reduce 
the theoretical uncertainties of the leading order (LO) predictions, that arise from the missing higher order 
quantum corrections through the renormalisation and factorisation scales.
This necessitates the calculation of the next-to-leading order (NLO) quantum effects
through Quantum Chromodynamics (QCD) radiative corrections to these SM observables at the hadron colliders.  
Presently, the phenomenological results for almost all physical processes of interest at the LHC are 
available at this accuracy due to the tremendous advancement in automatising the calculation of 
the virtual and real emission contributions. However, the situation becomes more involved in 
case of next-to-next-to leading order (NNLO) calculation due to a lot of technical difficulties. 
In order to improve the theoretical predictions in a consistent way, it is customary to take into account all 
the higher order contributions, which are important in the complementary kinematic regions of phase space 
corresponding to the phase space relevant for the fixed order evaluation. However, to act correctly on 
these effects invoke a lot of technical problems. In practice, it can be approximated via the 
parton shower (PS) algorithm, which not only gives a reasonable estimate of these effects in the 
collinear kinematic regions of the phase space, but also provides a very realistic final state configuration.  
In other words, parton level predictions have to be gone through such showering of multi partons 
and recombination of these partons into hadrons through a hadronisation
mechanism in order to compare them against the experimental data. Such predictions require careful 
matching of results at various orders to avoid double counting. Thus, NLO SM results 
supplemented with parton showering can provide a more reliable as well as realistic predictions 
that can serve in testing various BSM scenarios. Till now, there exists mainly two different algorithms 
incorporating the matching of a NLO calculation to parton showers, 
namely MC@NLO~\cite{MC@NLO} and POWHEG~\cite{powheg}. We shall adopt the MC@NLO algorithm here, which has 
already been implemented and completely automated in {\sc a}MC@NLO~\cite{amc@nlo}.

In this article, we revisit the three photon production process at the LHC at NLO in QCD and present study the 
consequences of matching it with the parton shower. Triple-photon production provides a background to 
techni-pion production in association with a photon, where the techni-pion decays into a photon 
pair~\cite{triplephoton}. This process has already been studied at LO~\cite{lo}, as well as at NLO 
level~\cite{nlo} in QCD. We extend the analysis including the effect of parton shower to get a realistic estimate
of various kinematical distributions. We quantify the improvement in the predictions at small transverse momentum 
regions of the final state particles and the stabilisation of cross section against the variation of the 
factorisation and renormalisation scales.

This paper is organized as follows: in section~\ref{calc}, we have described the details of the calculation, 
mainly the virtual as well as the real emission contribution. The numerical results of the fixed order 
calculation together with the NLO+PS accurate results have been discussed in section~\ref{results} and finally,  
we conclude in section~\ref{conclusion}.

\section{Calculational Details}
\label{calc}
LO (${\cal O}(\alpha^3$)) contributions to the production of three photons at the LHC 
come from quark anti-quark annihilation processes.  
At NLO ${\cal O}(\alpha^3 \alpha_s)$ in QCD,
we encounter virtual as well as real emission contributions resulting from 
an additional parton, namely quark or anti-quark or gluon.  
Virtual amplitudes are already
at ${\cal O}(\alpha^{3/2} \alpha_s)$, hence only the interference 
of them with the LO Born amplitudes will contribute to the NLO level.
The real emission processes at NLO level come from two types of processes
namely gluon emissions from the LO processes and 
scattering of a quark (anti-quark) and a gluon producing three photons along with a 
quark (anti-quark). The ultra-violet (UV) divergences coming from the
virtual contributions and the infra-red (IR) divergences originated
from the virtual as well as real emission contributions, need to
be removed through the addition of proper counter terms. The resulting IR-safe 
parton level cross section up to NLO can be written as, 
\begin{eqnarray}
d\hat \sigma_{ab}^{NLO}&=& \int dPS_{3\gamma} ~
S(\{p\}_{1,5})
~d\hat \sigma_{ab}^{(0)} 
\nonumber\\
&&+{\alpha_s(\mu_R) \over 4 \pi} \Bigg[\int dPS_{3\gamma}
~S(\{p\}_{1,5})
~d\hat \sigma_{ab}^{V,(1)}  
+\int dPS_{3\gamma}
~S(\{p\}_{1,5})
~d\hat \sigma_{ab}^{CT,(1)}  
\nonumber\\
&&+\int dPS_{3\gamma+parton}
~S(\{p\}_{1,6})
~d\hat \sigma_{ab}^{R,(1)}  
+\int dPS_{3\gamma}
~S(\{p\}_{1,5})
~d\hat \sigma_{ab}^{MF,(1)}  \Bigg]
\label{signlo}
\end{eqnarray}
The first term is the  Born contribution; $dPS_{3 \gamma}$ is the phase space measure
of the three photon final states and $S(\{p\}_{1,m})$ is the observable function which
depends on the kinematic variables through the momenta of the external particles {\it i.e.}, $p_1, p_2, \dots, p_m$.  
The second term corresponds to virtual corrections to the Born process. They are often divergent 
when the loop momentum becomes very large and these UV divergences are first regularised and then renormalised 
using the counter terms given in the third term. The fourth term represents the real emission contributions 
at the NLO level come from parton emissions from the initial and/or final state partons. 
Due to massless quarks, anti-quarks and gluons participating in the hard
processes, both virtual and real emission contributions encounter soft and 
collinear divergences. The divergences coming from soft gluons and from collinear partons 
in the final state of the real emission processes get cancelled with those coming from the virtual 
processes. The remaining collinear divergences from the initial states are removed 
by adding mass counter terms given in the last term of eq.~\ref{signlo}.
The details of obtaining UV renormalised virtual contributions are discussed in the next section. 
The real emission contributions and the corresponding mass counter terms are obtained with the 
help of M{\sc{ad}}FKS~\cite{madfks}, 
a set of routines available in the \aMCatNLO~\cite{amc@nlo}, which along with our in-house FORTRAN routines for 
calculating virtual contributions, can provide results on an event-by-event basis in terms of four momenta of all 
the particles involved in the scattering process and we use them to obtain the observables that we require to study. 
In the following sub-sections, we sketch a systematic outline of the complete computational procedure. 

\subsection{Virtual contribution}
\label{virt}
Virtual contribution comes from the interference between the Born diagrams and 
the one loop corrected virtual diagrams. 
The number of virtual diagrams to order $\alpha^{3/2} \alpha_s$ for the three 
photon production is forty eight. Up to permutations of the final state photons, 
we find 1 pentagon diagram, 2 box diagrams, 3 triangle diagrams and 
2 bubble diagrams. We have used QGRAF~\cite{qgraf} to generate both LO and NLO amplitudes. 
It generates the symbolic description of the Feynman diagrams in terms of propagators and vertices. 
We have written a FORM~\cite{form} code, which translates the output of QGRAF into
a suitable format, that can be used for further symbolic manipulations. 
We have supplied Feynman rules, identities for Dirac gamma matrices, equations of motion 
through this code and performed various simplifications at the amplitude level. The loop integrals 
are regulated using dimensional regularisation. Both Lorentz contractions and Dirac gamma matrix 
simplifications are done in $n=4+\varepsilon$ space-time dimensions. Both UV and IR divergences 
appear as poles in $\varepsilon$ and they have been calculated using $\overline{\mbox{MS}}$ scheme. 
Writing the virtual contribution in the following way:
\beq
\sum_{\rm col}   \sum_{\rm spin} 
 {\cal M}^{V,\rm (1)} \left ( {\cal M}^{(0)}
\right)^{*}
= 
\sum_{\Gamma}\left[\sum_{\rm col}\sum_{\rm spin}
{\cal M}^{V,(\Gamma)} \left ( {\cal M}^{(0)} \right)^{*}\right] \quad ,
\eeq
where ${\cal M}^{(0)}$ is the born amplitude and 
${\cal M}^{V,(\Gamma)}$s' are the distinct topologies of virtual
diagrams, we compute only one particular topology and then
the permutations of photon momenta and their polarisations gave us
the remaining contributions.  

The reduction of tensor integrals to scalar ones in $n$ dimensions is done using the standard procedure 
{\it a\!\!\textasciigrave ~la} Passarino-Veltman~\cite{PV_reduction}. 
The tensor integrals that appear at one loop level are of the form
\begin{eqnarray}
I_n^{\mu_1 \cdot \cdot \cdot \mu_m} = \int {d^n l \over (2 \pi)^n}
{l^{\mu_1} \cdot \cdot \cdot l^{\mu_m} \over 
((l-q_1)^2+i \epsilon) \cdot \cdot \cdot ((l-q_n)^2+i \epsilon)} \quad ,
\end{eqnarray}
where 
\begin{eqnarray}
q_1=p_1,\,\, q_2=p_1+p_2, \,\,\,\,\cdot \cdot \cdot,\,\,\,\, q_n= \sum_{i=1}^n p_i \quad .
\end{eqnarray}
One can decompose the above tensor integral in terms
of scalar coefficients as follows:
\begin{eqnarray}
I_n^{\mu_1 \cdot \cdot \cdot \mu_n} = \sum_{i_1,\cdot \cdot \cdot,i_m}^n
q_{i_1}^{[\mu_1} \cdot \cdot \cdot q_{i_m}^{\mu_m]} F_{i_1 \cdot \cdot \cdot
i_m}^{(n)}
+\sum_{i_3, \cdot \cdot \cdot,i_m}^n  g^{[\mu_1 \mu_2} q_{i_3}^{\mu_3}
\cdot \cdot \cdot q_{i_m}^{\mu_m]}  F_{00 i_3 \cdot \cdot \cdot
i_m}^{(n)}   \quad ,
\end{eqnarray}
where the square bracket implies the non-equivalent symmetrisation by giving the full set of 
non-equivalent permutations.
We have written a FORM code for the purpose of doing this tensor reduction and expressed
the virtual contributions in terms of these scalar co-efficients.
As described in~\cite{devy}, these co-efficients are related to the scalar integrals in different 
space-time dimensions in the following way, 
\begin{eqnarray}
I_{n, i_1, i_2, \cdot \cdot \cdot }^{[2 i], s_1, s_2, \cdot \cdot \cdot}
= \int {d^{n+2 i}l \over (2 \pi)^{n+2 i} }
\prod_{r=1}^n {1  \over ((l-q_r)^2+i \epsilon)^{1+\delta_{ri_1} +
\delta_{ri_2} + \cdot \cdot \cdot -\delta_{rs_1}-\delta_{rs_2} -\cdot \cdot \cdot}} \quad ,
\end{eqnarray}
where $I_{n, i_1, i_2, \cdot \cdot \cdot }^{[2 i], s_1, s_2, \cdot \cdot \cdot}$ is a generalized 
scalar integral in shifted space-time dimension. These integrals in the shifted dimensions 
can be expressed in terms of integrals in $n$ dimensions using the dimensional recurrence 
relations discussed in~\cite{tarasov}. In this approach, inverse Gram determinants that result from 
the recurrence relations, often spoil the numerical stability of the integral. There exists a handful of solutions to this problem 
in the literature~\cite{neerven, oldenborgh, campbell, binoth:2000, campanario:2011, giele,denner, binoth:2005, diakonidis}.  
Recently, an elegant approach has been put forward in~\cite{fleischer}, where the authors have found 
signed minor algebraic relation, which avoids the appearance of inverse Gram determinants and thereby 
introducing a set of higher dimensional scalar integrals to cope with the small Gram determinants. 
These higher dimensional scalar integrals have been evaluated numerically after employing a series 
expansion in the small Gram region.
This whole algorithm has been implemented in the numerical package, named PJFry~\cite{{pjfry thesis},{pjfry package}},
which we use to evaluate numerically the scalar co-efficients of the tensor integral for every phase space point 
in $n$ dimensions. PJFry reduction library uses QCDLoop~\cite{qcdloop} and OneLOop~\cite{oneloop} to evaluate
the scalar integrals in $4$ dimensions. In order to validate our
FORM codes, namely those ones that perform conversion of output of 
QGRAF to FORM readable symbolic expressions, reduction of tensor integrals 
to scalar coefficients and also to validate FORTRAN routines, which evaluate the virtual 
contributions numerically using PJFry, we re-calculated the virtual corrections of the di-photon production 
process in both SM and BSM to order $\alpha_s$. We compared our results thoroughly against the results presented
in~\cite{{kumar},{diph nlops}} and found an excellent agreement between these two. Using our FORM codes and 
FORTRAN routines along with the publicly available packages, {\it viz.} QGRAF, PJFry, QCDLoop and OneLOop, we have evaluated the 
virtual contributions to the three photon production process at ${\cal O}(\alpha_s)$ level.
We find that after UV renormalisation, the IR poles namely
double and single poles in $\varepsilon$ are in accordance with
the expectation. We express the virtual contribution of the 
three photon production in a form suitable for further analysis as follows:
\begin{eqnarray}
d\hat \sigma^{V,(1)}_{q \overline q} = \frac{\alpha_s}{2\pi} {1 \over \Gamma(1+{\varepsilon \over 2})}
\Bigg({s \over 4 \pi \mu_R^2}\Bigg)^{\varepsilon\over 2} C_F \Bigg(- {8\over \varepsilon^2} + {6 \over\varepsilon} \Bigg) 
d\hat \sigma^{(0)}_{q \overline q} + d\hat \sigma^{V,(1),fin}_{q \overline q}(\mu_R) \quad ,
\end{eqnarray}
where $\alpha_s$ is the strong coupling evaluated at the the renormalisation scale $\mu_R$, $s$ is the partonic center of mass energy 
and the colour factor is: $C_F=4/3$ for $SU(3)$. $d\hat \sigma_{q \overline q}^{(0)}$ 
comes from the colour-linked Born amplitude ${\cal M}_{q \overline q}^{(0)}$, whereas $d\hat \sigma^{V,(1),fin}_{q \overline q}$ denotes the finite 
virtual contribution that has been computed numerically. Note that the IR poles in $\varepsilon$ are in agreement with the 
universal behaviour of soft and collinear partons.

\subsection{Real emission contribution}
\label{realemission}
Real emission contributions come from gluon emission from the Born processes 
as well as from the scattering of a quark/anti-quark and a gluon producing a quark/anti-quark and 
three photons. We use \aMCatNLO~\cite{amc@nlo} framework not only to compute these 
contributions along with the mass factorisation terms required to remove the initial state 
collinear singularities, but also to obtain the NLO results matched with PS. Within \aMCatNLO, 
the stand-alone package M{\sc ad}G{\sc raph}~\cite{madgraph} generates all the required matrix 
elements both at LO as well as at NLO level. 
As already discussed in the previous sub-section \ref{virt}, we have prepared a set of external codes to deal with 
the virtual correction part and made an interface to implement it within M{\sc ad}FKS~\cite{madfks}, which 
separates out the soft and collinear configurations in the real 
emission processes using the FKS subtraction scheme~\cite{fks} and provides
IR-divergent and IR-safe contributions separately along with the mass factorisation
terms, that take part in removing the initial state collinear singularities coming from the virtual
and the real emission processes. In the FKS subtraction scheme, the phase space is partitioned in such 
a way that each partition contains at most one soft and one collinear divergences. This is done by 
introducing a set of positive-definite $S_{ij}$ functions, where the $S_{ij}$s' are chosen in such a way that 
they vanish in all singular limits not related to: ({\sf i}) a particle $i$ becoming soft, ({\sf ii}) particles $i$ and $j$ 
becoming collinear, obeying the restriction that the sum over all such pairs must be equal to identity. 
This ensures that each term of the sum is finite throughout the phase space, except when the energy of particle $i$ goes to zero or particles $i$ 
and $j$ become collinear. Now, after finding out the exact position of the divergences for a given partition, the 
generalized plus distribution is used to regulate them. All these steps are systematically automated in M{\sc ad}FKS 
within the M{\sc ad}G{\sc raph}5 environment. 
We have explicitly checked the cancellation of the soft and collinear
divergences among the virtual, real and mass factorisation terms at different regions of the phase space thereby 
confirming the perfect implementation of all the above mentioned external inputs within the \aMCatNLO\ framework. 
The events, that are generated using \aMCatNLO, also include the Monte Carlo 
counter terms to take care of the MC@NLO matching and thereby preventing the 
occurrence of any double counting at the time 
of matching to PS. These events are then showered by 
HERWIG~\cite{herwig}, PYTHIA~\cite{pythia} parton shower to get the realistic events. 

Photons are produced not only at the partonic level, but also through the fragmentation of 
partons into photons and a jet of hadrons can often be collinear to them. 
This necessitates the inclusion of non-perturbative fragmentation functions. 
At NLO level, the QED collinear divergence can arise when one of the final state parton becomes collinear to a photon. 
This can be factorised in a universal manner and then removed by adding counter terms, which renormalise the 
fragmentation functions, thereby bringing in a scale dependence at the partonic cross sections through the 
fragmentation functions, which is known as fragmentation scale. 
An alternate isolation criteria has been proposed in \cite{frixione}, using
which one can obtain an observable in which fragmentation contribution is
minimised and at the same time, the IR safety of that observable is guaranteed.
We call it Frixione isolation here after and use this isolation for our analysis.  It works in the following way: define a
cone centered around each photon with a 
radius $R$ in the rapidity-azimuthal angle ($\eta-\phi$) plane, where $R=\sqrt{(\eta-\eta_{\gamma})^2+(\phi-\phi_{\gamma})^2}$. 
Now, it is demanded that the sum of hadronic transverse energy $H(R)$ inside any concentric circle of radius $R < R_{\gamma}$ would be 
less than an amount given by the function $H(R)_{max}$.  This function can be chosen in such a way that 
lesser and lesser hadronic energy is allowed as we move closer to a given photon. 
Because of the fact that $H(R)$ goes to zero as $R \rightarrow 0$, the partons that are collinear
to photon are removed while the soft partons are kept intact thereby guaranteeing the QCD IR safety.
For our analysis, we have taken the following canonical choice for $H(R)_{max}$, {\it i.e.},
\begin{equation}
H(R)_{max} = \epsilon_{\gamma}\ {E_{T}^{\gamma}}\ \Bigg(\frac{1-\cos R}{1-\cos R_{\gamma}}\Bigg)^n \quad ,
\text{\quad } \, 
\label{eq:frixione}
\end{equation}
where $E_{T}^{\gamma}$ is the transverse energy of the photon and $R_{\gamma}$, $\epsilon_{\gamma}$, $n$ 
are three parameters that are to be set while applying this isolation 
criteria.\footnote{Effects of photon frangmentation and different isolation
prescriptions have very recently been studied \cite{campbell_williams}}

\begin{figure}
\centerline{ 
\includegraphics[width=7.5cm]{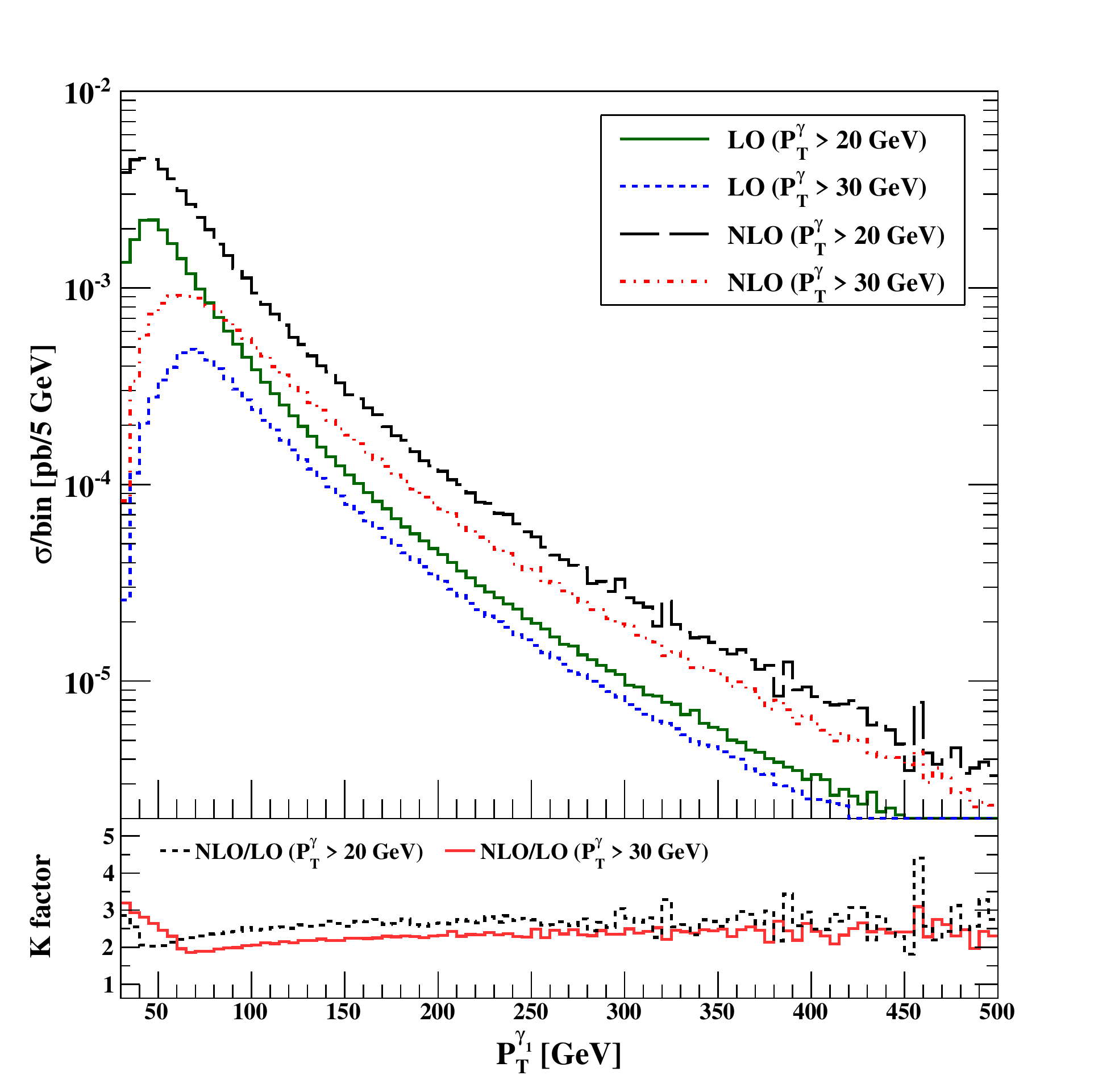}
\includegraphics[width=7.5cm]{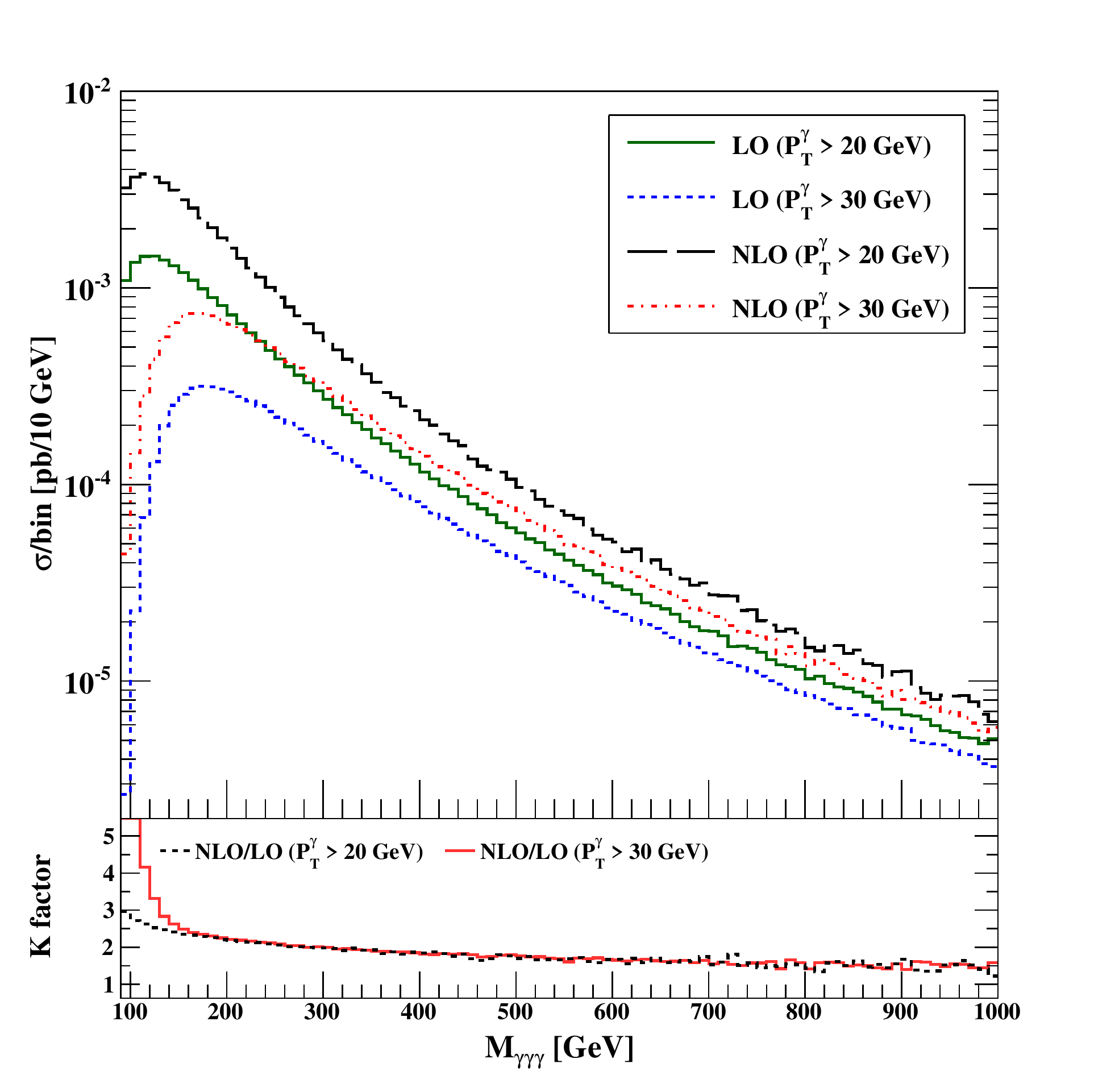}
}
\caption{\label{fig_pt} Transverse momentum distribution of the hardest photon $P_T^{\gamma_1}$ 
(left panel) and invariant mass distribution $M_{\gamma \gamma \gamma}$ 
of the three photon (right panel) for the fixed order NLO and LO. 
}
\end{figure}
\section{Numerical Results}
\label{results} 
In this section, we present the results for various kinematic distributions relevant to the 
production of three photon in SM at the LHC with the center-of-mass energy $\sqrt{S}=14$ TeV. 
Here we list the input parameters used for the whole computation:
\begin{align}
&M_Z = 91.188 \ \mathrm{GeV} , &&\alpha^{-1}_{em} = 132.507 , \nonumber\\
&G_F = 1.16639 \times 10^{-5} \ \mathrm{GeV}^{-2}  \; . &&  
\label{eq:ew}
\end{align}
\\
These values of $\alpha_{em}$, $G_F$ and $M_Z$ ensure that the mass of the W-boson ($M_W = 80.419$ GeV) 
and the value of $\sin^2{\theta_W}$ ($\sin^2{\theta_W} = 0.222$) remain closer to the experimental values. 
We have considered massless quarks with five flavours ($n_{f}=5$) throughout our calculation. 
\begin{figure}
\centerline{ 
\includegraphics[width=7.5cm]{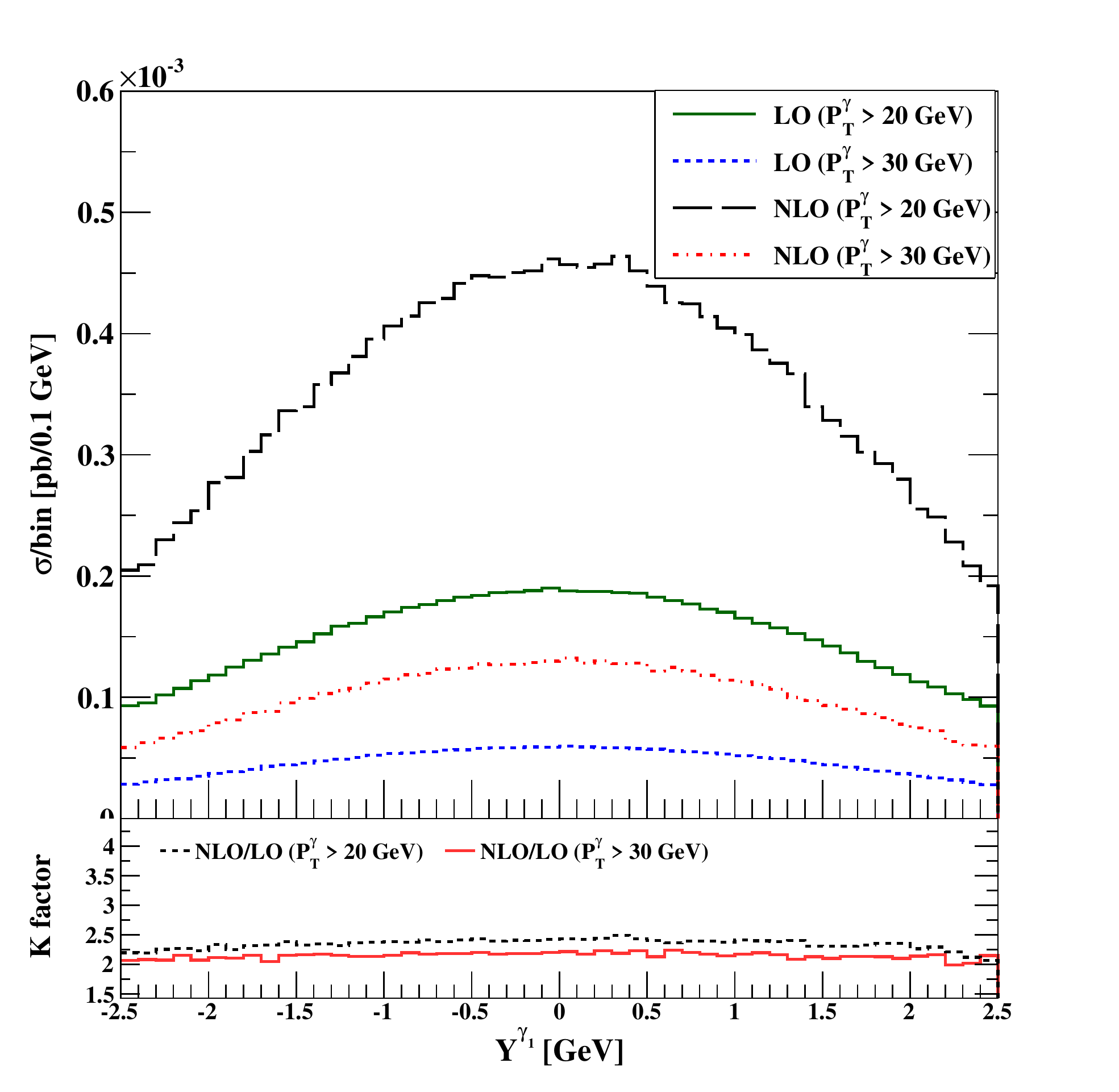}
\includegraphics[width=7.5cm]{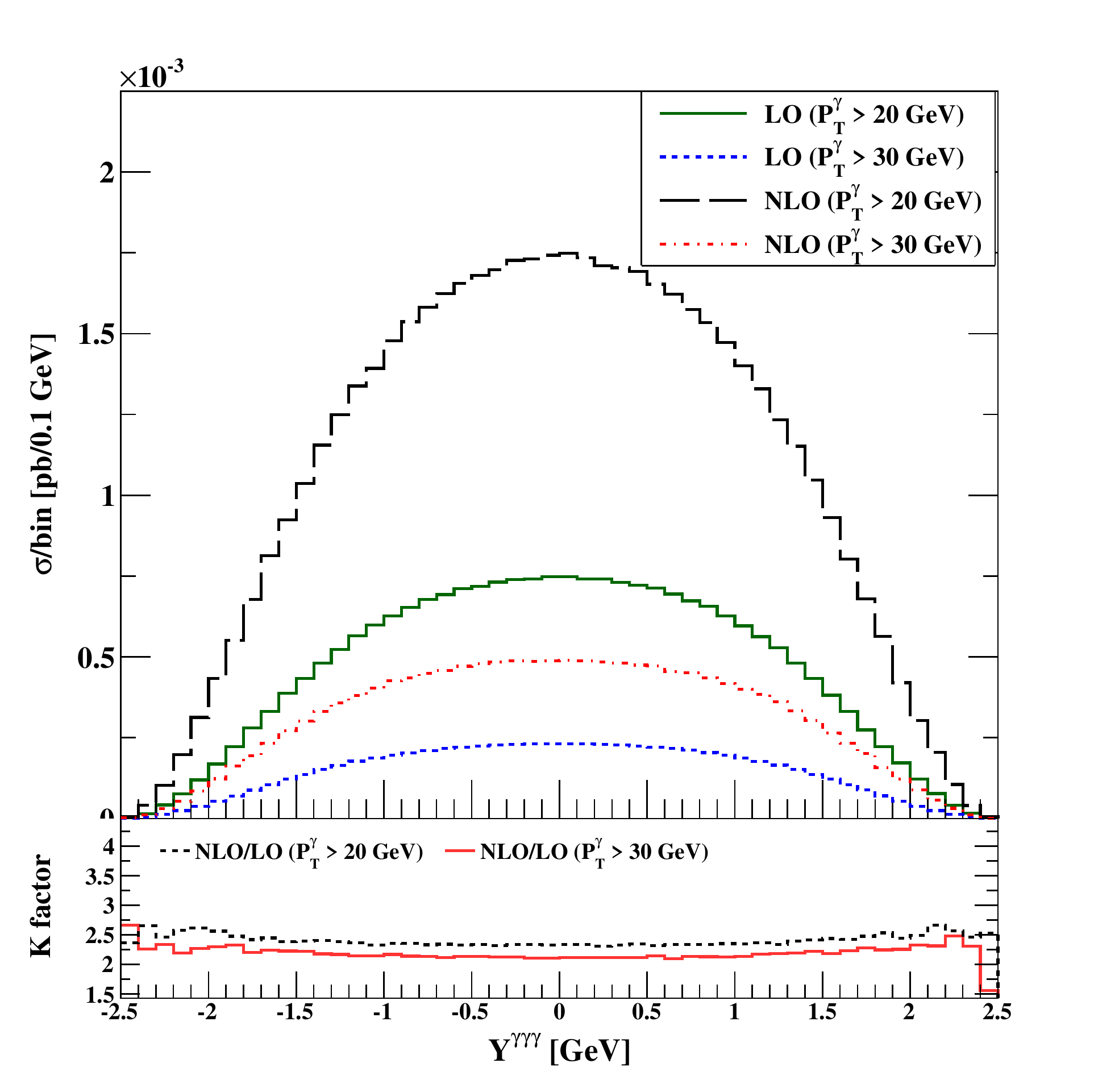}
}
\caption{\label{fig_rap} Rapidity distribution of the hardest photon $Y^{\gamma_1}$ 
(left panel) and  the three photon system $Y^{\gamma \gamma \gamma}$ (right panel) for the fixed order NLO and LO.
}
\end{figure}
In our present study, we have used MSTW2008(N)LO parton 
distribution function with errors estimated at 68$\%$ CL for the (N)LO and it also sets the value of the 
strong coupling $\alpha_{s}(M_{Z})$ at (N)LO in QCD. 
The factorisation scale ($\mu_F$) and the renormalisation scale ($\mu_R$) are set equal to a central 
scale, which is the invariant mass of the three photon final states {\it i.e.}, 
$\mu_F=\mu_R=M_{\gamma\gamma\gamma} \equiv\sqrt{(P_{\gamma_1} + P_{\gamma_2} + P_{\gamma_3})^2}$ .

For the fixed order (N)LO calculation, we have taken the following choices of cuts: 
rapidity of each photon $|\eta^{\gamma}| < 2.5$, separation between any two photons in the ($\eta-\phi$) plane $\Delta R^{\gamma\gamma} > 0.4 $, 
where $\Delta R^{\gamma\gamma} = \sqrt{(\Delta \eta)^2 + (\Delta \phi)^2}$. 
In addition, we have studied a variety of differential distributions applying two types of cuts on the transverse momentum of 
each photon {\it i.e.}, $P_T^{\gamma} > 20$ GeV and $P_T^{\gamma} > 30$ GeV in the 
fixed order analysis. 
Unless stated otherwise, we consider $P_T^{\gamma} > 30$ GeV as the generic choice of cut on photons transverse momenta. 
Parameters involved in the Frixione isolation are set as: $R_{\gamma}=0.7$, $\epsilon_\gamma =1$ and $n=2$. 

\begin{table}[h!]
  \begin{center}
    \begin{tabular*}{0.99\textwidth}{@{\extracolsep{\fill}}|l|cc|cc|cc|}
     \hline
      ~~LHC 
       &\multicolumn{1}{c}{LO [pb]} &
       &\multicolumn{1}{c}{NLO [pb]} &
       & K-factor & \\ \hline
      ~~$P_T^{\gamma} > 20$ GeV
       & \phantom{0} $2.257\times 10^{-2}$\phantom{00} &
       & \phantom{}$5.336\times 10^{-2}$\phantom{00} &
       & 2.36 & \\       
      ~~$P_T^{\gamma} > 30$ GeV
       & \phantom{0}$7.050\times 10^{-3}$\phantom{0} &
       & \phantom{}$1.519\times 10^{-2}$\phantom{00} &
       & 2.16 & \\ \hline
    \end{tabular*}
    \caption[]{Total cross sections for the 3-photon production at the LHC.                     
     The results are shown for two different cuts at LO, NLO, and the associated K-factor. 
     Relative statistical errors of the Monte Carlo are below $10^{-5}$.}
    \label{tab:LHC}
  \end{center}
\end{table}
\subsection{Fixed order Analysis}
\label{foresults}
\begin{figure}
\centerline{ 
\includegraphics[width=8cm]{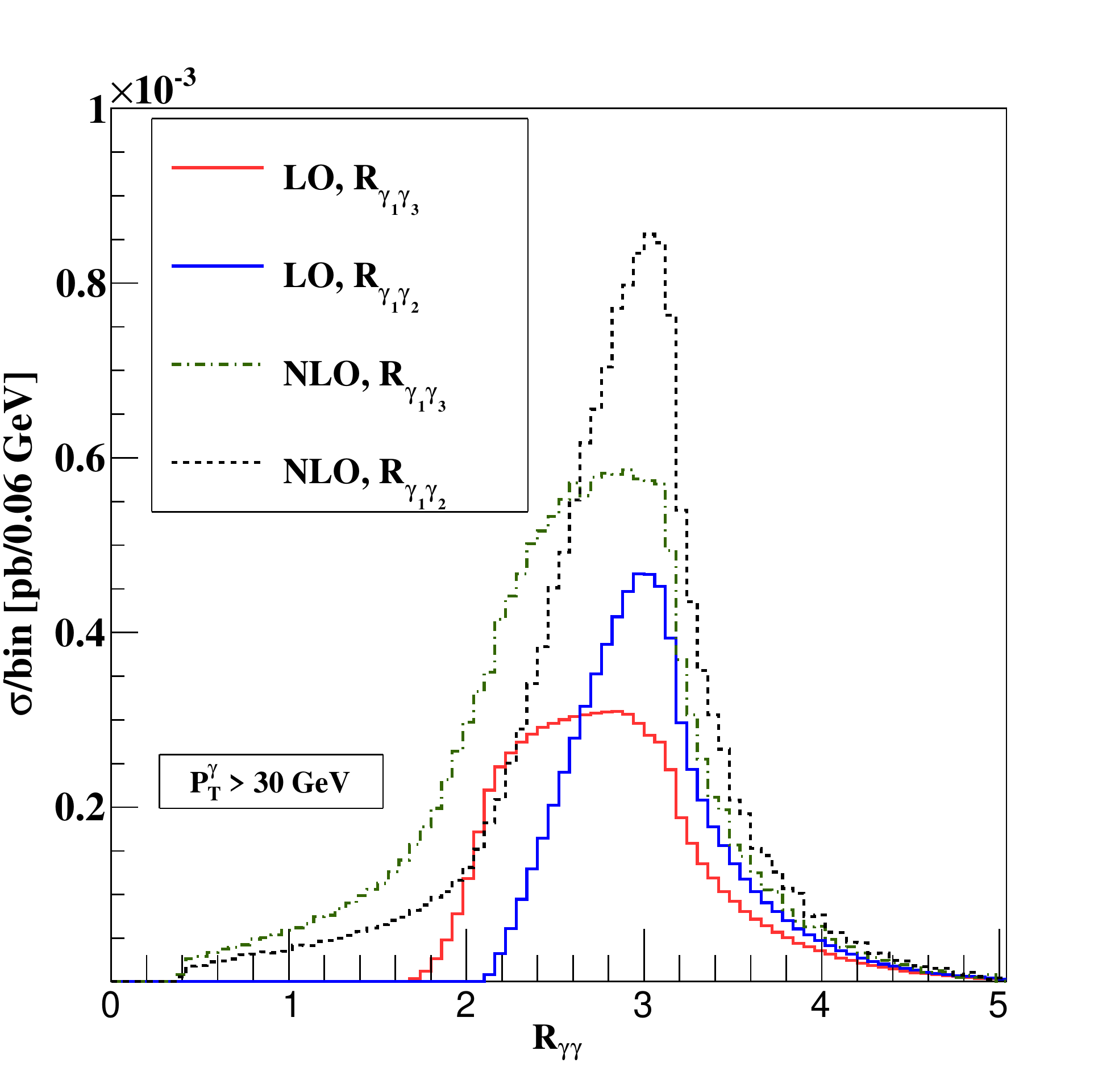}
}
\caption{\label{fig_sepfo} 
Separation of the softer photons ($\gamma_{2},\gamma_{3}$) in comparison with the hardest one ($\gamma_{1}$)
at LO and NLO. 
}
\end{figure}
In table~\ref{tab:LHC}, we have shown the results of total cross sections for fixed order LO and NLO using the central choice of 
$\mu_F$ and $\mu_R$ for two different $P_T^{\gamma}$ cuts. 
To begin with, we present some distributions of few selective kinematical variables at fixed order LO and NLO. Photons are 
ordered according to their transverse momentum. The hardest photon with maximum transverse momentum is denoted by $\gamma_1$.  
Like wise, $\gamma_2$ represents the second hardest photon and the softest photon is labelled as $\gamma_3$. 
In fig.~\ref{fig_pt}, we have shown transverse momentum distribution of $\gamma_1$ at LO and NLO in the left panel 
and in the right panel, distribution of invariant mass of the three photon system has been plotted. The lower insets 
show the bin-by-bin distribution of the K-factor for the corresponding observable. We find that, for low transverse 
momentum, the K-factor is large as it is due to the fact that the recoil against the extra parton helps to fulfil 
the transverse momentum cut, which was not possible at LO. The left panel of the fig.~\ref{fig_rap} shows the 
distribution of the rapidity of the hardest photon, whereas the rapidity of the three photon system is shown in the right panel. 
The distribution of the K-factor is shown for the corresponding variables in the lower insets. Unlike fig.~\ref{fig_pt}, 
the K-factors in fig.~\ref{fig_rap} appear to be mostly steady indicating the affinity of these observables towards the 
photons having fairly high transverse momenta. 
\begin{figure}
\centerline{ 
\includegraphics[width=7.5cm]{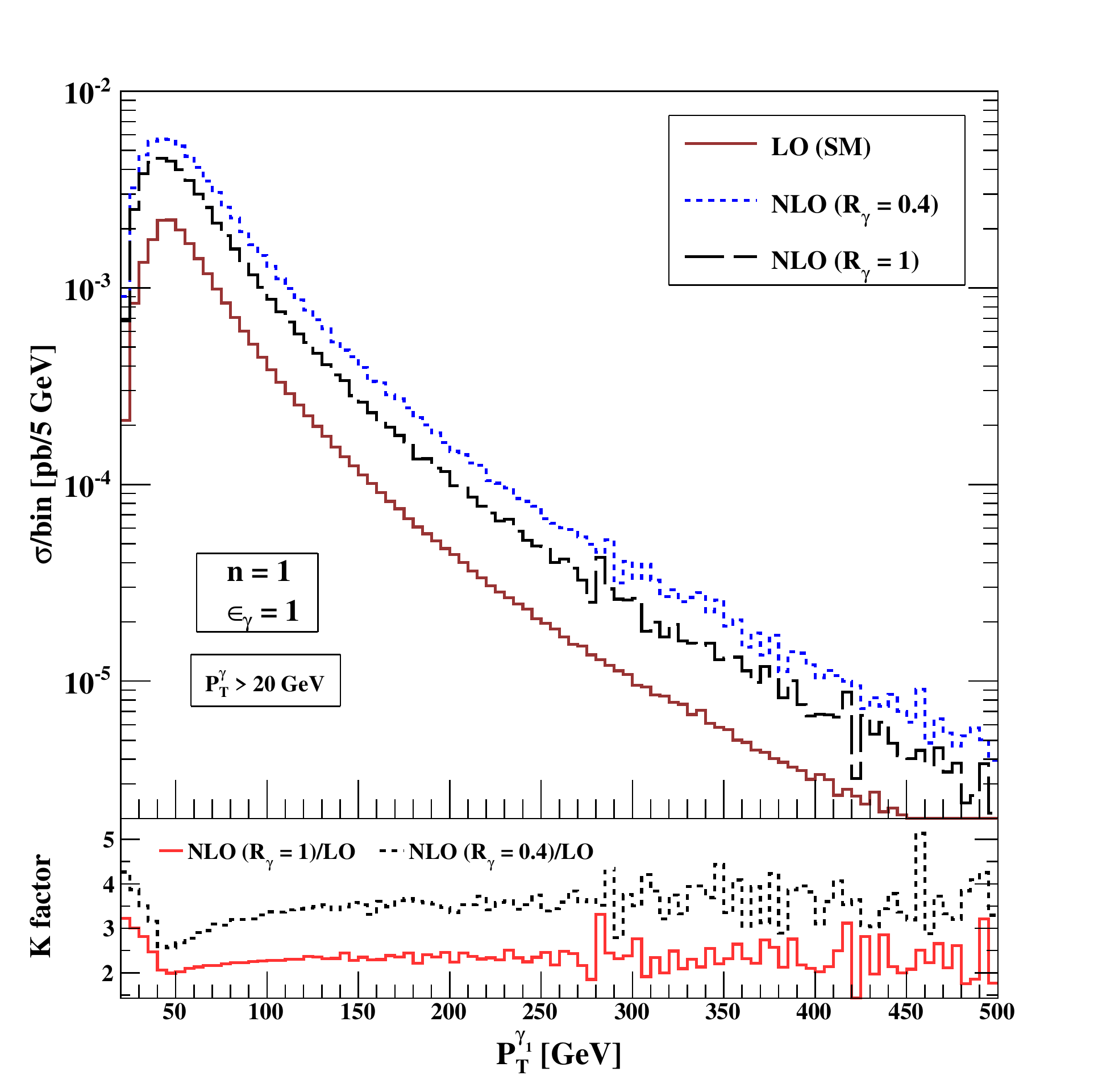}
\includegraphics[width=7.5cm]{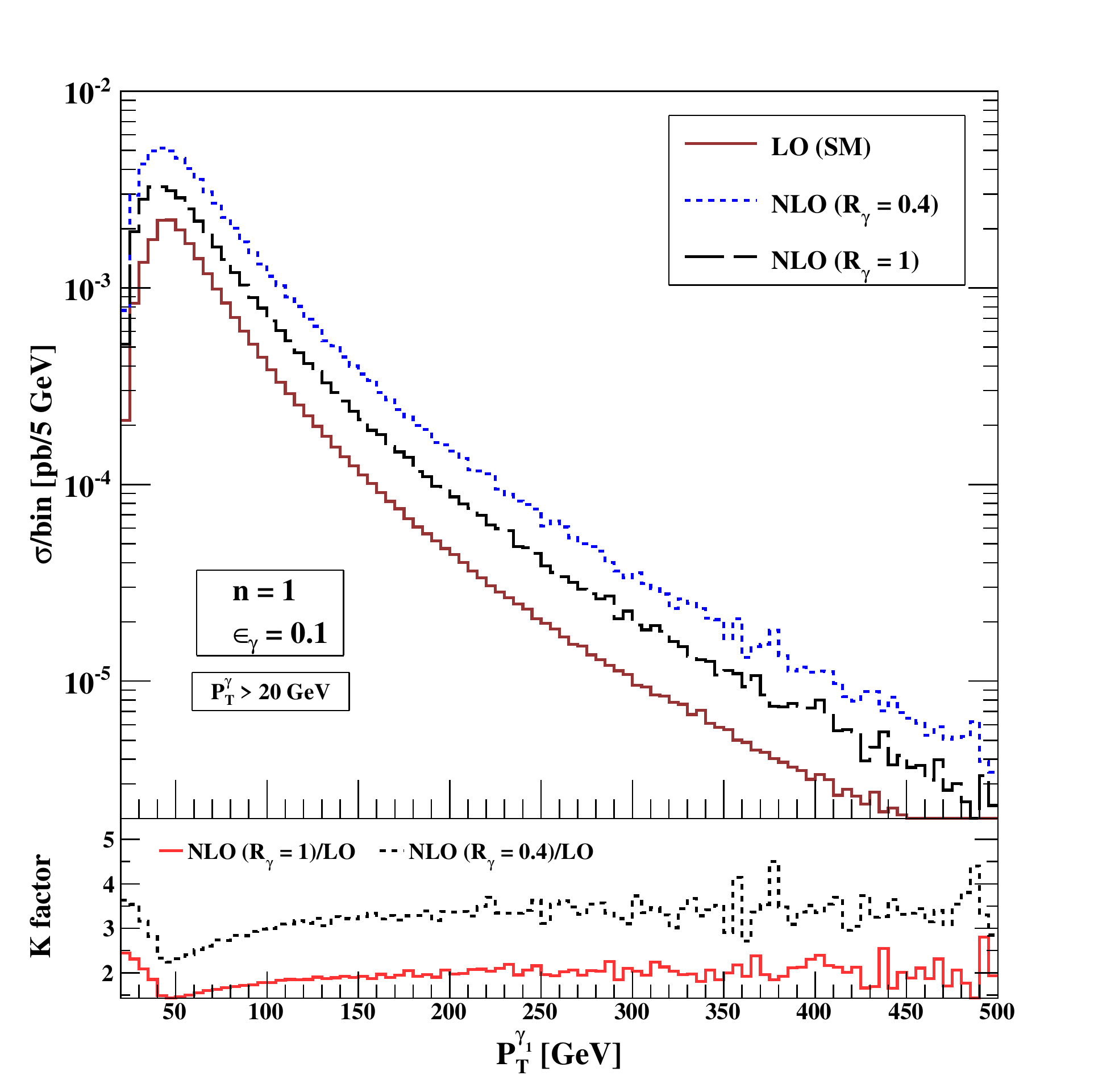}
}
\caption{\label{fig_isolr} Transverse momentum distribution of the hardest 
photon ($P_T^{\gamma_1}$) with $R_{\gamma}$ variation for a fixed value of n=1 and $\epsilon_{\gamma}$ =1 (left panel) and 
for another fixed value of n=1 and $\epsilon_{\gamma}$ =0.1 (right panel).
}
\end{figure}

\color{black}
\begin{table}[h!]
  \begin{center}
\begin{tabular}{cc|c|c|c|c}
\cline{1-5}
\multicolumn{1}{| c| }{$R_{\gamma}$}& n & \multicolumn{3}{| c| }{$\sigma_{NLO}$\ [pb]} \\ \cline{3-5}

\multicolumn{1}{| c| }{}& & $\epsilon_{\gamma}=1$ & $\epsilon_{\gamma}=0.5$ & $\epsilon_{\gamma}=0.1$ \\ 

\cline{1-5}

\multicolumn{1}{ |c| }{\multirow{2}{*}{0.4} } &
\multicolumn{1}{ |c| }{1} & $6.896 \times 10^{-2}$ & $6.550 \times 10^{-2}$ & $6.154 \times 10^{-2}$ &     \\ \cline{2-5}
\multicolumn{1}{ |c  }{}                        &
\multicolumn{1}{ |c| }{2} & $6.489 \times 10^{-2}$ & $6.291 \times 10^{-2}$ & $6.045 \times 10^{-2}$ &     \\ \cline{1-5}
\multicolumn{1}{ |c  }{\multirow{2}{*}{1} } &
\multicolumn{1}{ |c| }{1} & $5.090 \times 10^{-2}$ & $4.620 \times 10^{-2}$ & $3.825 \times 10^{-2}$ & \\ \cline{2-5}
\multicolumn{1}{ |c  }{}                        &
\multicolumn{1}{ |c| }{2} & $4.454 \times 10^{-2}$ & $4.110 \times 10^{-2}$ & $3.462 \times 10^{-2}$ & \\ \cline{1-5}
\end{tabular}
    \caption[]{Total cross sections for the 3-photon production at the LHC for various Frixione isolation parameters. We have taken $p_T^{\gamma} > 20$ GeV at NLO.}
    \label{tab:FI}
  \end{center}
\end{table}

All the above distributions show a substantial effect of radiative corrections on this process. This is mainly 
because of the inclusion of new subprocesses at the NLO, as quark-gluon subprocesses begin to contribute at this order and 
due to the enhancement in the phase space. In fig.~\ref{fig_sepfo}, we have plotted the separation between the ordered 
photons in the ($\eta-\phi$) plane obeying the selection cut: $\Delta R^{\gamma_i\gamma_j} > 0.4$, where $i,j=1,2,3$. 
We have checked the rapidity differences between these photons are quite small. Therefore, the peaks arising in these distributions 
near the angle $\pi$ (180$\mathring{}$), suggest that the emitted photons are mostly back-to-back. The hardest photon $\gamma_1$ is separated form 
the softest one {\it i.e.}, $\gamma_3$, by at least $\Delta R^{\gamma_1\gamma_3} = 1.6$ at LO, whereas at NLO they can be very 
close as permitted by the selection cut due to the emission of an extra radiation at this level. 
\\
\begin{figure}
\centerline{ 
\includegraphics[width=8cm]{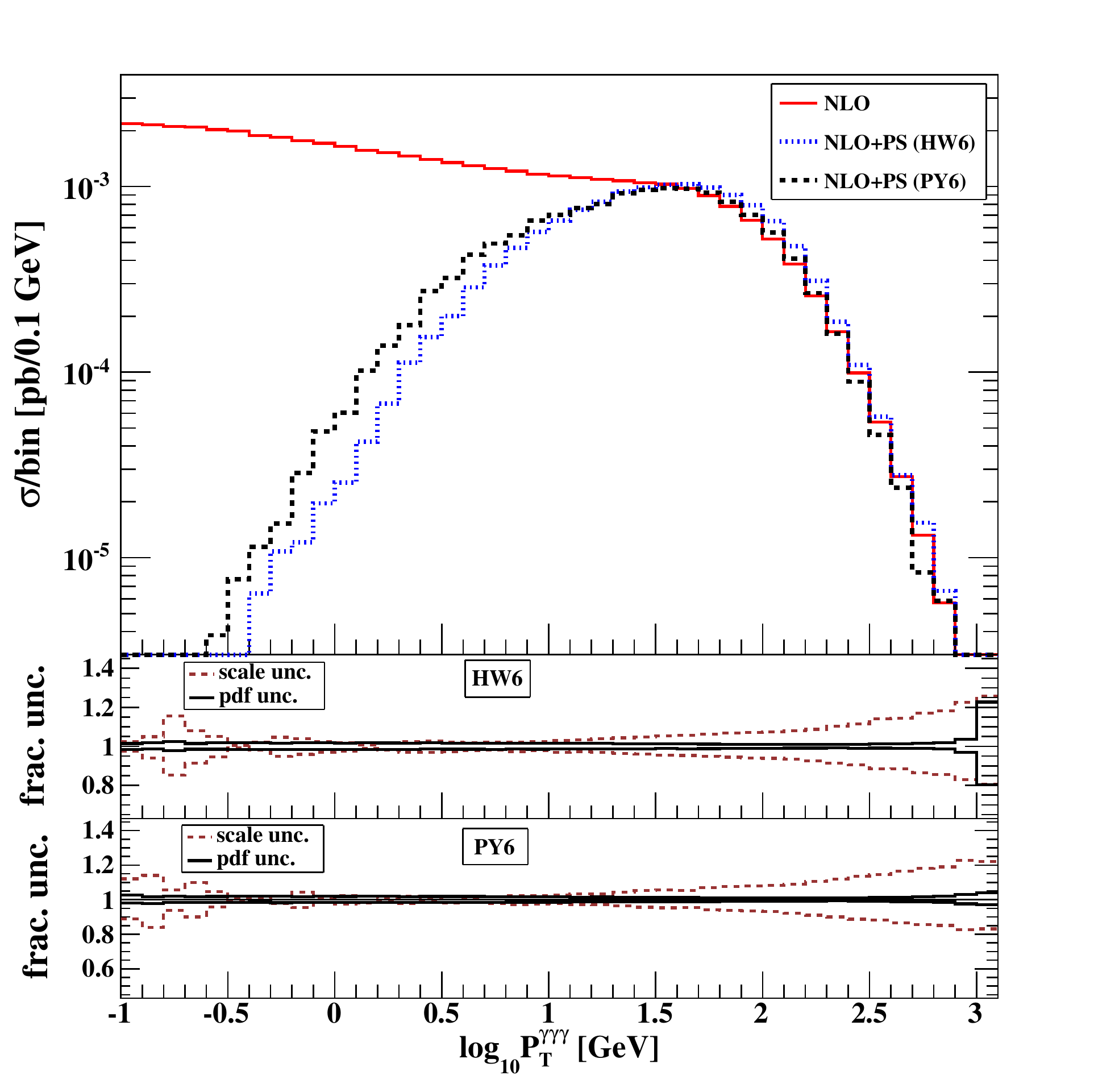}
}
\caption{\label{fig_logpt} Three photon transverse momentum distribution $P_T^{\gamma\gamma\gamma}$ for the fixed order NLO and NLO+PS. 
}
\end{figure}
Besides, we have checked the effect of variation of Frixione isolation parameters {\it i.e.}, $R_{\gamma}$, $\epsilon_{\gamma}$ 
and n. 
Though Frixione isolation has no effect on the LO cross section, the dependency of the NLO cross section on these isolation parameters is shown in table~\ref{tab:FI}. From eq.~(\ref{eq:frixione}), it is evident that the NLO cross-section increases when $R_{\gamma}$ decreases 
and it also increases with increasing $\epsilon_{\gamma}$. 
In fig.~\ref{fig_isolr}, we have shown the transverse momentum distribution of the hardest photon by varying the 
value of $R_{\gamma}$ (left panel) from $0.4$ to $1$ for a fixed value of $n=1$ and $\epsilon_{\gamma}$ =1, where the 
K-factors vary from $3.06$ to $2.26$. The right panel shows the same distribution for a fixed value 
of $n=1$ and $\epsilon_{\gamma}=0.1$ and in this case,  
K-factors vary from $2.72$ to $1.69$. 
It is evident from table~\ref{tab:FI}, as well as from fig.~\ref{fig_isolr} that, for a fixed choice of $n$ value, 
the NLO cross-section is large for $R_{\gamma}=0.4$ and $\epsilon_{\gamma}=1$, 
whereas it becomes much smaller for $R_{\gamma}=1$ and $\epsilon_{\gamma}=0.1$ indicating the fact that 
smaller $R_{\gamma}$ increases the cross-section when $\epsilon_{\gamma}$ is larger~\cite{fabio_diphotonjet}. 
It is also clear from table~\ref{tab:FI}, that the effect of varying $\epsilon_{\gamma}$, keeping n and $R_{\gamma}$ 
fixed, is quite minimal. Similar studies with changing the value $n=2$, provide same kind of distributions analogous to fig.~\ref{fig_isolr}. 
\subsection{Discussion on NLO+PS}
\label{nlopsresults}
In this section, we compare the fixed order NLO result with the NLO results matched with PS (NLO+PS) with two different 
showering algorithm, namely HW6 and PY6. 
For the showering purpose, parton level events are generated using very loose cuts: 
$P_T^{\gamma} > 15$ GeV, $|\eta^{\gamma}| < 2.7$, $\Delta R^{\gamma\gamma} > 0.3$ with the following Frixione isolation parameters:  
$R_{\gamma}=0.4$, $\epsilon_\gamma =1$ and $n=2$. 
We have explicitly checked that the events thus produced, remain unbiased in total rates
and differential distributions after showering and hadronisation for this choice of 
kinematical cuts and Frixione isolation parameters. These events are then showered with
HERWIG6 (HW6) and PYTHIA6 (PY6) and we have imposed the same set of analysis cuts that
we used in the fixed order analysis along with the generic $P_T^{\gamma}$ cut on the
transverse momentum of the photon at the time of showering. 

The scale dependencies of the results are calculated by varying $\mu_F$ and $\mu_R$ independently around the
central value $\mu_F = \mu_R = M_{\gamma\gamma\gamma}$ via the following assignment: $\mu_F=\xi_F~M_{\gamma\gamma\gamma}$ 
and $\mu_R=\xi_R~M_{\gamma\gamma\gamma}$, where $\xi_F$ and $\xi_R$ are varied between the range [1/2,2] independently. 
Various ratios of $\mu_F$, $\mu_R$ and $M_{\gamma\gamma \gamma}$ that appear as arguments of logarithms 
in the perturbative expansion to NLO are within the range [1/2,2]. 
The scale uncertainty band is the envelope of the results obtained by varying this $\xi_F$ and $\xi_R$ within this range~\cite{diph nlops}. 
The PDF uncertainties 
are estimated with the Hessian method, as given by the MSTW~\cite{mstw} collaboration. We have plotted fractional uncertainty, 
which is defined as the ratio of the variation about the central value divided by the central value, being a 
good indicator of the uncertainties. These uncertainty bands can be generated automatically at the time of 
parton level event generation by storing additional 
information, sufficient to determine via a reweighting technique, at no extra CPU cost within the \aMCatNLO\ framework as described in~\cite{scale}.

 We have shown $\log_{10} P_T^{\gamma\gamma\gamma}$ distribution for HW6 and PY6 together the fixed order NLO result, 
in fig.~\ref{fig_logpt}. It is clear that at low $P_T^{\gamma\gamma\gamma}$ values, NLO+PS (for both HW6 and PY6) 
result shows the effect of all order resummation of the large logarithms, hereby suppressing the cross section leading 
to a meaningful value, while the fixed order NLO result diverges for $P_T^{\gamma\gamma \gamma} \to 0$. 
At low $P_T^{\gamma\gamma\gamma}$, PY6 result is different from the HW6 result as the soft and collinear 
emissions constituting the parton shower are treated differently. PYTHIA generates more softer spectra than HERWIG 
in this region and as a result of this, these two showers show different behaviour as expected~\cite{pythia amcatnlo}. 
At high $P_T^{\gamma\gamma\gamma}$, the NLO fixed order and NLO+PS (for both HW6 and PY6) results are in agreement 
as in this region, the hard emissions are dominant and they are correctly described by the NLO hard cross section. 
In the middle and lower insets of fig.~\ref{fig_logpt}, we have presented the fractional scale and PDF uncertainties 
of the NLO+PS result for HW6 and PY6 respectively which increase with increasing $P_T^{\gamma\gamma\gamma}$~\cite{pythia amcatnlo}.
We do not find any significant differences in case of studying fractional uncertainties using these two different showers. Therefore, 
in the rest of the figures, we present the fractional uncertainty plots only for HW6. 

\begin{figure}
\centerline{ 
\includegraphics[width=7.5cm]{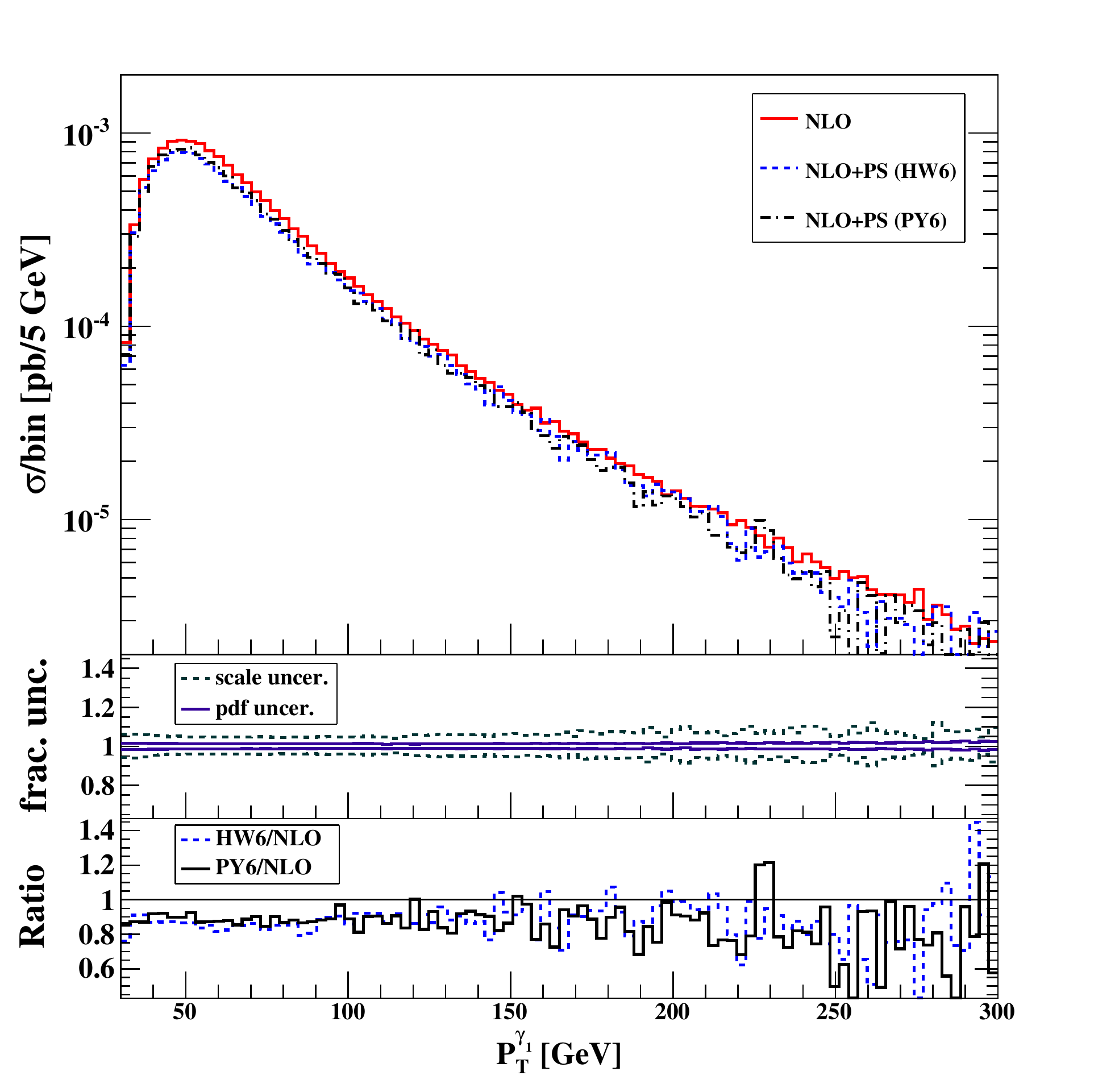}
\includegraphics[width=7.5cm]{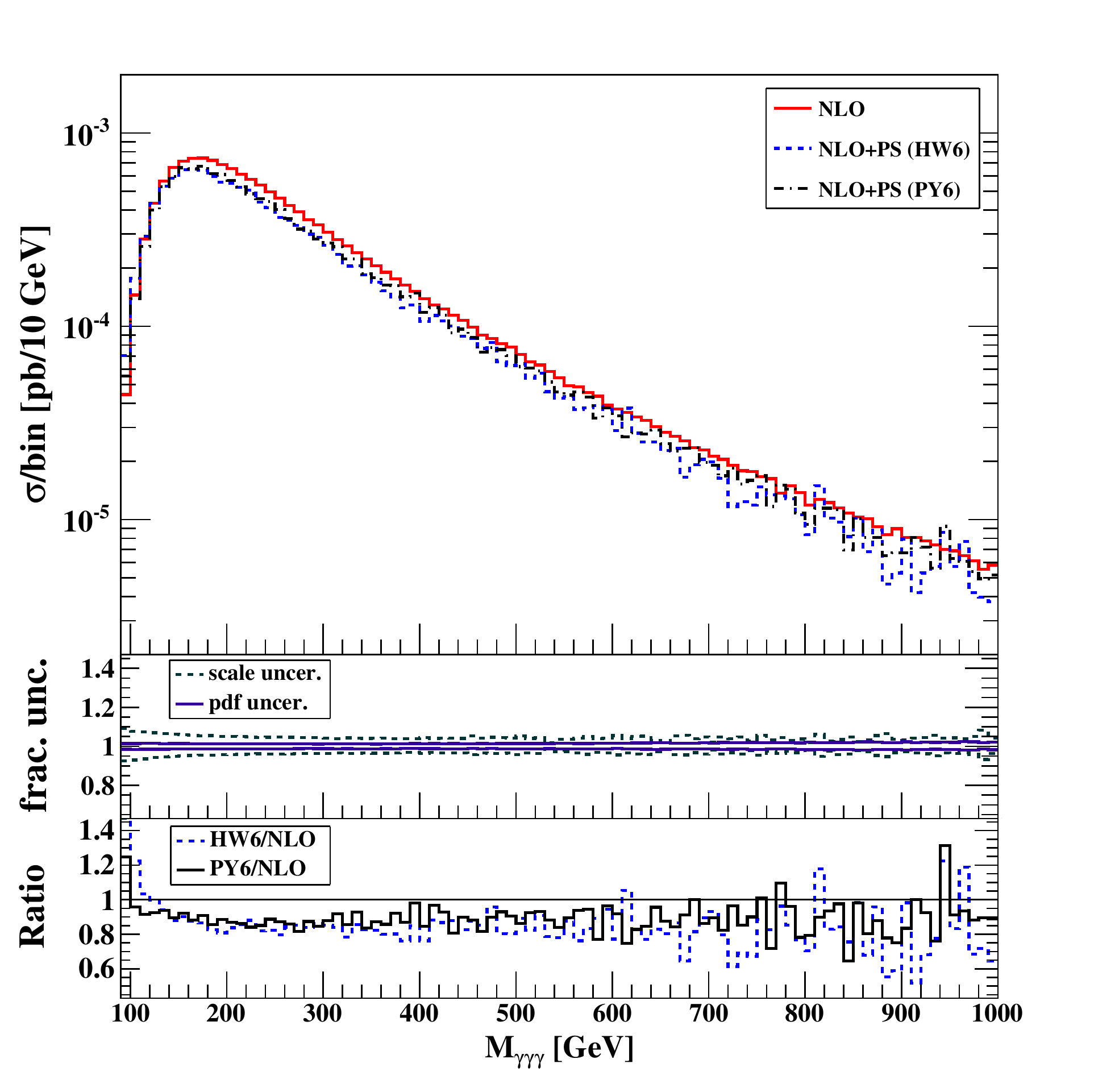}
}
\caption{\label{fig_a1pt} Transverse momentum distribution $P_T^{\gamma_1}$ of the hardest photon (left panel) 
and invariant mass distribution $M_{\gamma\gamma\gamma}$ of the three photon system (right panel) for the fixed order NLO and NLO+PS.
}
\end{figure}
We now present the results for various kinematical distributions to NLO accuracy, matched with PS (labelled as NLO+PS), 
for both HW6 and PY6 with the specified analysis cuts. We have adopted a consistent pattern for all the rest of the distributions. 
In each case, within the main frame, three curves corresponding to the distributions in fixed order NLO (solid red) and 
NLO+PS using HW6 (dashed blue) and NLO+PS using PY6 (dotted black) are shown. The middle inset shows fractional scale 
uncertainty (dashed cyan) and fractional pdf uncertainty (solid violet), while the lower inset shows the ratio between 
NLO+PS and NLO for HW6 (dashed blue) and for PY6 (solid black).
%
\begin{figure}
\centerline{ 
\includegraphics[width=8cm]{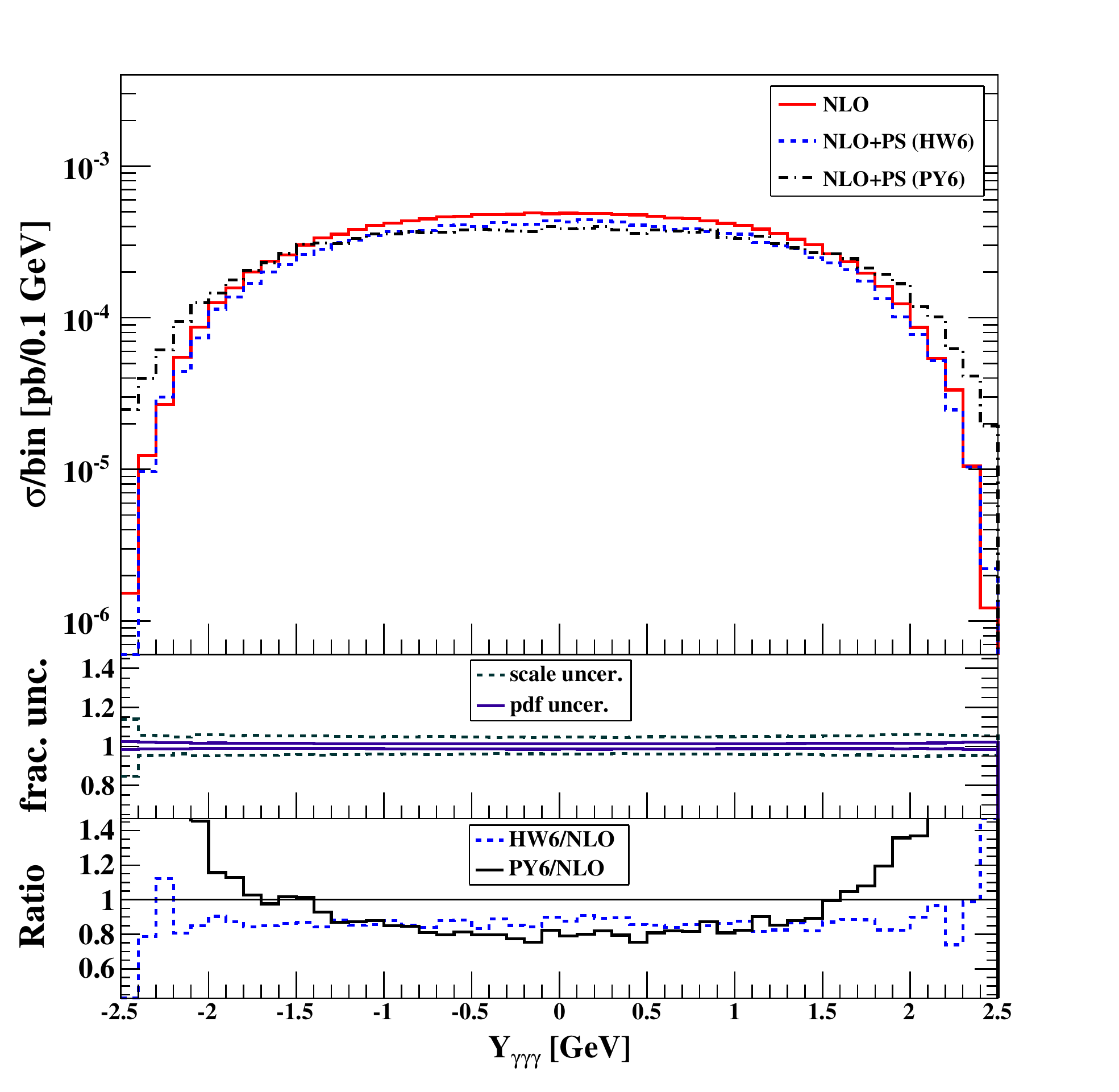}
}
\caption{\label{fig_rapps} Three photon rapidity distribution for the fixed order NLO and NLO+PS. 
}
\end{figure}
In the left panel of fig.~\ref{fig_a1pt}, we have shown the plots for transverse momentum distribution of the 
hardest photon and the right panel shows the distribution of the invariant mass distribution of the three photons.
We do not find much difference in the results of two showers HW6 and PY6. 
In both distributions, NLO results are very little larger than the NLO+PS results and this is due to the fact that 
the QCD radiation becomes softer when we demand all the photons to satisfy a high $P_T^{\gamma}$ cut ({\it i.e.} $P_T^{\gamma} > 30$ GeV) 
and damping of PDFs at large Bjorken $x$ values further subdue its effect at the parton level, whereas \aMCatNLO\ produces more events with 
hard and central jets resulting in the suppression of these distributions after showering. In fig~\ref{fig_rapps}, we have depicted the plot showing 
rapidity distribution of the three photon system and as expected, we observe that the NLO result is slightly harder than the NLO+PS result. 
However, the ratio in the lower inset shows that PY6 generated events give larger contribution than HW6 in the large rapidity region indicating that 
PY6 produces significantly large number of radiations than HW6 in the full kinematically available phase space. 
\section{Conclusion}
\label{conclusion}
Precise and realistic predictions of both signal and background processes at hadron colliders 
are now possible due to tremendous developments in the computational methods and the 
availability of the state of the art computational tools. We have used packages, namely QGRAF, PJFry, \aMCatNLO~ to 
study the three photon production process at the NLO level in QCD for the LHC taking into account the parton shower effects 
and realistic experimental cuts. In addition, we have developed some codes that build the interfaces among these different 
analytical and numerical tools. 
We have plotted different kinematic observables and discussed the consequences of showering the fixed order NLO results 
with two different showering algorithm HERWIG and PYTHIA. We have also discussed the effects of scanning over the Frixione isolation 
parameters on the NLO cross section. We find our predictions are less sensitive to scale uncertainties and choice 
of PDFs and hence more suited for direct comparison with the data from the experiments.

\section*{Acknowledgments}
We would like to thank Rikkert Frederix for many helpful discussions and acknowledge Valery Yundin for his help in running PJFry package. 
The work of MKM and VR has been partially supported by funding from Regional Center for Accelerator-based Particle Physics (RECAPP), 
Department of Atomic Energy, Govt. of India. PM and SS would like to acknowledge HPC cluster computing facility of Theory Division, SINP.



\begin{thebibliography}{99}

\bibitem{wishlist}
J.M. Campbell, J.W. Huston, W.J. Stirling, Rep. Prog. Phys. 70 (2007) 89,
The SM and NLO Multileg Working Group: Summary report, arXiv:1003.1241, The SM and NLO Multileg and SM MC Working Groups,: Summary Report, arXiv:1203.6803.

\bibitem{MC@NLO} 
  S.~Frixione and B.~R.~Webber,
  JHEP {\bf 0206}, 029 (2002)
  [hep-ph/0204244].
  
\bibitem{powheg} 
  P.~Nason,
  JHEP {\bf 0411}, 040 (2004)
  [hep-ph/0409146];\\
  S.~Frixione, P.~Nason and C.~Oleari,
  JHEP {\bf 0711}, 070 (2007)
  [arXiv:0709.2092 [hep-ph]].

\bibitem{amc@nlo}
R. Frederix, S. Frixione, V. Hirschi, F. Maltoni, R. Pittau and P.Torrielli,
Phys. Lett. B701 (2011) 427;
R. Frederix, S. Frixione, V. Hirschi, F. Maltoni, R.Pittau and P. Torrielli,
JHEP 09 (2011) 061.
  
\bibitem{triplephoton}
  A.~Zerwekh, C.~Dib, R.~Rosenfeld,
  Phys.\ Lett.\  {\bf B549}, 154-158 (2002).
  [hep-ph/0207270].
  
\bibitem{lo}
M. Golden, S.R. Sharpe, Nucl. Phys. B261 (1985) 217,
V. Barger, T. Han, Phys. Lett. B212 (1988) 117.

\bibitem{nlo}
G. Bozzi, F. Campanario, M. Rauch, D. Zeppenfeld, Phys. Rev. D84 (2011) 074028.

\bibitem{madfks}
R. Frederix, S. Frixione, F. Maltoni and T. Stelzer, JHEP 10 (2009) 003.

\bibitem{qgraf}
P. Nogueira, Journal of Computational Physics 105 (1993) 279.

\bibitem{form}
M.Tentyukov and J.A.M. Vermaseren, hep-ph/0702279.

\bibitem{PV_reduction}
L. M. Brown and R. P. Feynman, Phys. Rev. 85 (1952) 231–244,
G. ’t Hooft and M. J. G. Veltman, Nucl. Phys. B153 (1979) 365–401,
G. Passarino and M. J. G. Veltman,Nucl. Phys. B160 (1979) 151.

\bibitem{devy}
A. I. Davydychev, Phys. Lett. B263 (1991) 107.

\bibitem{tarasov}
O.V. Tarasov, Phys. Rev. D54 (1996) 6479, 
J. Fleischer, F.Jegerlehner, and O.V. Tarasov, Nucl. Phys. B566 (2000) 423.

\bibitem{neerven}
W. L. van Neerven and J. A. M. Vermaseren, Phys. Lett. B137 (1984) 241

\bibitem{oldenborgh}
G. J. van Oldenborgh and J. A. M. Vermaseren, Z. Phys. C46 (1990) 425.

\bibitem{campbell}
J. M. Campbell, E. W. N. Glover, and D. J. Miller, Nucl. Phys. B498 (1997) 397.

\bibitem{binoth:2000}
T. Binoth, J. P. Guillet, and G. Heinrich, Nucl. Phys. B572 (2000) 361.

\bibitem{campanario:2011}
F. Campanario, JHEP 1110, 070 (2011), F. Campanario, Q. Li, M. Rauch, M. Spira, JHEP 1306, 069 (2013), F. Campanario, H. Czy ̇ , J. Gluza, M. Gunia, T. Riemann, G. Rodrigo, V. Yundin, JHEP 1402, 114 (2014)


\bibitem{giele}
W. T. Giele and E. W. N. Glover, JHEP 04 (2004)029.

\bibitem{denner}
A. Denner and S. Dittmaier, Nucl. Phys. B734 (2006) 62–115.

\bibitem{binoth:2005}
T. Binoth, J. P. Guillet, G. Heinrich, E. Pilon, and C. Schubert, JHEP 10 (2005) 015.

\bibitem{diakonidis}
T. Diakonidis, J. Fleischer, J. Gluza, K. Kajda, T. Riemann, and J. B. Tausk, Phys. Rev. D80 (2009) 036003.

\bibitem{fleischer}
J. Fleischer and T. Riemann, Phys. Rev. D83 (2011) 073004.

\bibitem{pjfry thesis}
V. Yundin, Ph.D thesis, Humboldt-Universitat zu Berlin, 2012,
\newblock \href{http://edoc.hu-berlin.de/docviews/abstract.php?id=39163}{\path{http://edoc.hu-berlin.de/docviews/abstract.php?id=39163}}.

\bibitem{pjfry package}
\href{https://github.com/Vayu/PJFry/}{\path{https://github.com/Vayu/PJFry/}}.

\bibitem{qcdloop}
R.K. Ellis and G. Zanderighi, JHEP 02 (2008) 002.

\bibitem{oneloop}
A. van Hameren, Comput. Phys. Commun. 182 (2011) 2427.

\bibitem{kumar}
M.~C.~Kumar, P.~Mathews, V.~Ravindran and A.~Tripathi,
  Nucl.\ Phys.\ B {\bf 818} (2009) 28.
M.~C.~Kumar, P.~Mathews, V.~Ravindran and A.~Tripathi,
Phys.\ Lett.\ B {\bf 672} (2009) 45

\bibitem{diph nlops}
R. Frederix, M. K. Mandal, P. Mathews, V. Ravindran, S. Seth, P. Torrielli, M. Zaro, JHEP 1212 (2012) 102;

\bibitem{madgraph}
T. Stelzer and W.F. Long, Comput. Phys. Commun. 81 (1994) 357,
J. Alwall et al., JHEP 0709 (2007) 028.

%

\bibitem{fks}
S. Frixione, Z. Kunszt and A. Signer, Nucl. Phys. B467 (1996) 399,
S. Frixione, Nucl. Phys. B507 (1997) 295

\bibitem{herwig}
G. Marchenini et al., HERWIG, Comput. Phys. Commun. 67 (1992) 465,
G. Corcella et al., HERWIG 6.5, JHEP 01 (2001) 010,
G. Corcella et al., HERWIG 6.5 release note, hep-ph/0210213.

\bibitem{pythia}
T. Sjostrand, S. Mrenna and P.Z. Skands, JHEP 05 (2006) 026.

\bibitem{frixione}
S. Frixione, Phys.Lett. B429 (1998) 369.

\bibitem{campbell_williams}
J.\ M.\ Campbell, C.\ Williams, arXiv:1403.2641.

\bibitem{fabio_diphotonjet} 
Vittorio Del Duca, Fabio Maltoni, Zoltan Nagy, Zoltan Trocsanyi, JHEP 04 (2003) 059

\bibitem{mstw}
A.\ D.\ Martin, W.\ J.\ Stirling, R.\ S.\ Thorne, and G.\ Watt,
Eur.\ Phys.\ J.\ C63 (2009) 189–285.

\bibitem{scale}
R.\ Frederix, S.\ Frixione, V.\ Hirschi, F.\ Maltoni, R.\ Pittau and P.\ Torrielli, 
JHEP 02 (2012) 099.

\bibitem{pythia amcatnlo}
Paolo Torrielli, Stefano Frixione, JHEP 04 (2010)110.

\end{thebibliography}
\end{document}